\documentstyle[11pt,aaspp4,flushrt]{article}

\def\longpagen[#1]{\setlength{\textheight}{#1}}
\newcommand{\normalpage}{\setlength{\textheight}{8.4in}}

\newcommand\gsim{\mathrel{\raise.3ex\hbox{$>$}\mkern-14mu
             \lower0.6ex\hbox{$\sim$}}}
\newcommand\lsim{\mathrel{\raise.3ex\hbox{$<$}\mkern-14mu
             \lower0.6ex\hbox{$\sim$}}}

\def\arcs{\ifmmode {^{\scriptscriptstyle\prime\prime}}
          \else $^{\scriptscriptstyle\prime\prime}$\fi}
\def\arcm{\ifmmode {^{\scriptscriptstyle\prime}}
          \else $^{\scriptscriptstyle\prime}$\fi}
\newdimen\sa  \newdimen\sb
\def\parcs{\sa=.07em \sb=.03em
     \ifmmode $\rlap{.}$^{\scriptscriptstyle\prime\kern -\sb\prime}$\kern -\sa$
     \else \rlap{.}$^{\scriptscriptstyle\prime\kern -\sb\prime}$\kern -\sa\fi}
\def\parcm{\sa=.08em \sb=.03em
     \ifmmode $\rlap{.}\kern\sa$^{\scriptscriptstyle\prime}$\kern-\sb$
     \else \rlap{.}\kern\sa$^{\scriptscriptstyle\prime}$\kern-\sb\fi}

\begin{document}

\title{CONSTRAINTS ON $H_0$ FROM THE CENTRAL VELOCITY DISPERSIONS OF LENS GALAXIES}
\author{Aaron J. Romanowsky and Christopher S. Kochanek}
\affil{Harvard-Smithsonian Center for Astrophysics,
       MS-10, 60 Garden Street,
       Cambridge MA 02138 \\
       Email: aromanowsky@cfa.harvard.edu}
\authoraddr{MS-10 \\ 
       60 Garden Street \\
       Cambridge MA 02138  \protect \\
       Email: aromanowsky@cfa.harvard.edu}
 
\authoremail{aromanowsky@cfa.harvard.edu}
\authoremail{ckochanek@cfa.harvard.edu}

\begin{abstract}
We employ Schwarzschild's method of orbit modeling to constrain the mass 
profiles of the central lens galaxies in Q0957+561 and PG 1115+080.
We combine the measured central projected stellar velocity dispersions
of these galaxies with the self-similar radial profiles of the rms velocity
and of the Gauss-Hermite moment $h_4$ observed in nearby galaxies for 
$0 \lsim R \lsim 2 R_{\rm eff}$.  For Q0957+561, we find a 16\% uncertainty 
in the galaxy mass, and 
formal 2-$\sigma$ limits on the Hubble constant of
$H_0 = 61^{+13}_{-15}$ km s$^{-1}$ Mpc$^{-1}$.  For PG 1115+080, we find 
that none of the viable lens models can be ruled out,
so that $H_0$ is not yet strongly constrained by this system.
\end{abstract}
\keywords{
galaxies: elliptical and lenticular, cD ---
galaxies: structure ---
galaxies: fundamental parameters ---
galaxies: kinematics and dynamics ---
gravitational lensing ---
quasars: individual (Q0957+561) ---
quasars: individual (PG 1115+080) ---
distance scale ---
dark matter
}

\section{INTRODUCTION}

\longpagen[8.52in]
The gravitational lens system Q0957+561 (Walsh, Carswell, \& Weymann 1979)
has been modeled extensively in an effort to determine the Hubble constant 
$H_0$ from measurements of the time delay between its two primary images.
Since the long-running dispute over the time delay measurement has been 
resolved in favor of the short delay (Schild \& Thomson 1997; Haarsma 
{\it et al}. 1997, 1998; Kundi\'{c} {\it et al}. 1997), the largest remaining 
uncertainty arises from the mass model.  The lens consists of a cluster 
with a large central elliptical galaxy G1.  The asymmetric radial positions
of the images accurately constrain the parameterized radial mass 
distribution of G1 (Grogin \& Narayan 1996, hereafter GN).  However, the 
presence of the cluster introduces a degeneracy in the overall mass 
normalization of G1 (Falco, Gorenstein, \& Shapiro 1985; Gorenstein, Falco,
\& Shapiro 1988), and thus in the determination of the Hubble constant
($H_0 \propto \sigma_0^2$, where $\sigma_0$ is a velocity dispersion 
characterizing the mass of G1).  Therefore, additional independent 
constraints are needed on the relative contributions of G1 and the cluster
to the $6\arcs$ image separation.  These can be obtained by inferring 
{\it the mass of the cluster} from cluster dynamics (Garrett, Walsh, \& 
Carswell 1992; Angonin-Willaime, Soucail, \& Vanderriest 1994), from hot 
intracluster gas X-ray emission (Chartas {\it et al}. 1995, 1998), or from 
the weak lensing of background galaxies (Dahle, Maddox, \& Lilje 1994; 
Fischer {\it et al}. 1997).  Alternatively, one can infer {\it the mass of 
G1} from stellar dynamical measurements.

\normalpage
The quadruple lens PG 1115+080 (Weymann {\it et al}. 1980) is the second 
system with a well-determined time delay (Schechter {\it et al}. 1997; 
Barkana 1997b).  The projected mass of the primary lens galaxy G inside the
ring of images can be precisely determined (unlike the case of Q0957+561, 
uncertainties in the mass distribution of nearby galaxies have only a minor
effect on the models).  However, the geometry of the system does not permit distinguishing between different mass profiles.  This has important 
consequences for the Hubble constant, since $H_0$ can vary by 40\%, 
depending on the mass model assumed (see Keeton \& Kochanek 1997; Courbin 
{\it et al}. 1997; Saha \& Williams 1997; Impey {\it et al}. 1998).
Stellar dynamical measurements of G may be helpful for breaking the 
degeneracy in the mass model, and thus in $H_0$.

Falco {\it et al}. (1997) measured the central projected stellar velocity 
dispersion of Q0957+561 G1 to be $\hat{\sigma}_{\rm p} = 279 \pm 12$ 
km s$^{-1}$, improving on an earlier measurement by Rhee (1991).
Similarly, Tonry (1998) measured the central dispersion of PG 1115+080 G
to be $\hat{\sigma}_{\rm p} = 281 \pm 25$ km s$^{-1}$.  However, the 
conversion of the measured $\hat{\sigma}_{\rm p}$ to $\sigma_0$ is subject 
to systematic uncertainties, which include the unknown anisotropy structure 
of the stellar orbits, the radial variation of the mass-to-light ratio, and
the ellipticity of the galaxy.  Previous galaxy models have arrived at a 
relatively small uncertainty in this conversion by making arbitrary 
simplifying assumptions.  Kochanek (1993, 1994) considered a singular 
isothermal mass model with constant anisotropy.  For Q0957+561, GN used 
these dynamical models, but limited them to be nearly isotropic, leading to
a 2\% systematic uncertainty in $\sigma_0^2$ (and thus in $H_0$).  Barkana 
(1997a) also assumed near-isotropic orbits, leading to an uncertainty in 
$H_0$ of 4\%, but he noted that allowing for more anisotropy gives an 
uncertainty of 14\%.  In fact, dynamical studies of nearby elliptical 
galaxies have clearly demonstrated that there is little basis for the 
assumption of isotropy, or even of constant anisotropy.

Binney \& Mamon (1982) and Tonry (1983) first illustrated that an 
elliptical galaxy's surface brightness and projected stellar velocity 
dispersion profiles, $I(R)$ and $\sigma_{\rm p}(R)$, could not determine 
both its mass distribution and its anisotropy profile.  Richstone \& 
Tremaine (1984) and Katz \& Richstone (1985) used orbit modeling methods to
demonstrate that the conversion from the dispersion profile 
$\sigma_{\rm p}(R)$ to a mass parameter $\sigma_0^2$ can be uncertain by an
order of magnitude.  Further theoretical studies (Dejonghe 1987; Merritt 
1987; Merrifield \& Kent 1990; Gerhard 1991; Dejonghe \& Merritt 1992; 
Merritt \& Saha 1993; Merritt 1993) demonstrated that {\it complete} 
knowledge of the stellar line-of-sight velocity distribution (LOSVD) 
${\cal L}(v_{\rm p},R)$ gives a unique solution for the two-integral 
distribution function (DF) $f(E,L)$ in a known spherical potential $\Phi$,
and may even strongly constrain an unknown $\Phi$.  The general efficacy of 
{\it incomplete} knowledge of ${\cal L}$ is less clear, but important 
constraints on $f$ and $\Phi$ could be further provided by large-radius 
measurements of higher-order velocity moments ({\it e.g.}, the 
Gauss-Hermite moments $h_l$ --- van der Marel \& Franx 1993; Gerhard 1993; 
see also Rix \& White 1992; Zhao \& Prada 1996).  Rix {\it et al}. (1997) 
and Gerhard {\it et al}. (1998) made dynamical fits to nearby galaxies
with higher-order moments ($h_3,h_4$) measured to $\sim 2.5 R_{\rm eff}$;
they determined the total mass to $\sim$  10-15\%, and ruled out a constant
mass-to-light ratio with $>99\%$ confidence.  For the galaxy NGC 2434, the 
solutions typically had nearly constant radial anisotropy 
$\beta(r) \equiv (1-v_\theta^2/v_r^2) \simeq 0.5$, while NGC 6703 showed an 
anisotropy rising from $\beta \sim 0.1$ at the center to $\sim 0.4$ near 
$R_{\rm eff}$.  {\it Thus the assumption of $\beta(r)=0$, or even of constant 
$\beta(r)$, is certainly unwarranted and probably incorrect. }

While our current knowledge of elliptical galaxies does not permit us to
make arbitrary assumptions about the anisotropy of the DF, we can impose 
the constraint that the unmeasured $\sigma_{\rm p}(R)$ and $h_4(R)$ 
profiles of these lens galaxies are similar to those of other galaxies
(provided they are universally homologous).  In this study we rigorously 
consider the utility of the central $\hat{\sigma}_{\rm p}$ measurement for 
determining $H_0$ from these two lens systems.  To ensure physically 
correct, robust results, we use a spherical orbit modeling method after 
Schwarzschild (1979), Richstone \& Tremaine (1984), and Rix {\it et al}. 
(1997) --- a fully general way to construct realistic models of a galaxy, 
given an assumed potential.  In \S 2 we review the observational 
constraints on the lens galaxies, and introduce constraints on their 
LOSVD profiles by demonstrating that the profiles of nearby elliptical 
galaxies are self-similar.  We describe our modeling method and demonstrate
it with a test-case problem in \S 3.  In \S 4, we report the range of model
solutions for Q0957+561, and discuss the implications for $H_0$.  We examine
the solutions for PG 1115+080 in \S 5.  In \S 6 we give our conclusions.

\normalpage
\section{OBSERVATIONAL CONSTRAINTS}

We first review the observational data for Q0957+561 G1 and PG 1115+080 G in
\S 2.1.  As discussed in \S 1, a simple measurement of the central 
projected stellar velocity dispersion $\hat{\sigma}_{\rm p}$ cannot 
strongly constrain the mass of a lens galaxy, necessitating further 
{\it a priori} constraints.  However, rather than making unjustified 
assumptions about the galaxy's anisotropy, we impose conditions on its 
observable properties by requiring it to have an LOSVD profile consistent 
with those of better-observed galaxies.  To this end, we extract ``mean 
profiles'' of $v_{\rm rms}(R)$ and $h_4(R)$, including their 
galaxy-to-galaxy scatter, from the observational data available for nearby 
elliptical galaxies\footnote{As introduced by van der Marel \& Franx (1993) 
and Gerhard (1993), ${\cal L}$ is parameterized by the Gauss-Hermite
moments, 
\begin{equation} 
h_l \equiv \sqrt{2}\frac{\gamma_0}{\hat{\gamma}}\int_{-\infty}^{\infty} {\cal L}(v_{\rm p}) e^{-\hat{w}^2/2}H_l(\hat{w})dv_{\rm p},
\end{equation}
where $\hat{w}=(v_{\rm p}-\hat{v}_{\rm p})/\hat{\sigma}_{\rm p}$, 
$\gamma_0$ is the line strength, 
$(\hat{\gamma},\hat{v}_{\rm p},\hat{\sigma}_{\rm p})$ are the coefficients
for the best Gaussian fit to ${\cal L}$, and $H_l(\hat{w})$ are the Hermite 
polynomials.  A perfectly Gaussian profile will have $h_l = \delta_{0 l}$
and $\hat{\sigma}_{\rm p} = \sigma_{\rm p}$.  For spherical systems, the 
fourth moment $h_4$ is a useful constraint on the DF; its value is 
typically positive (peaked LOSVD) if $f$ is radially anisotropic, and 
negative (flat-topped LOSVD) if tangentially anisotropic.  Strictly 
speaking, since we apply the data to non-rotating models, we should use the
Gauss-Hermite moment $z_4$ corresponding to the {\it even} part of the 
LOSVD (van der Marel {\it et al}. 1994 \S 5.1), but such data are not 
available, and the correction is probably small.}.  If there proves to be a
universal shape to these profiles ({\it i.e.}, there is little scatter 
between galaxies), we can use them as additional constraints (see \S 2.2).

\subsection{Observations of Q0957+561 and PG 1115+080}
Accurate observations of Q0957+561 G1 are difficult because of the nearby 
bright quasar image, but measurements have been made of the stellar surface
brightness profile $I(R)$ and the central velocity dispersion 
$\hat{\sigma}_{\rm p}$.  Bernstein, Tyson, \& Kochanek (1993) measured 
$I(a)$ along the major axis ($a=$ 2\parcs5-18\parcs3) in the $R$-band, and 
Bernstein {\it et al}. (1997) measured it at smaller radii ($a=$ 
0\parcs1-5\parcs7) in the $V$-band.  The galaxy is rather round, with an 
ellipticity increasing with radius ($\epsilon \simeq 0.1 , 0.2 , 0.4$ for 
$a \simeq 0\parcs1 , 1\parcs5 , 10\arcs$).  We combine the $R$ and $V$ 
data, assuming a simple offset of 1.34 magnitudes, and map the data to the 
intermediate radius $m\equiv \sqrt{a(1-\epsilon)}$, resulting in a profile 
$I(m)$ with $m=$ 0\parcs1-12\parcs8, where $R_{\rm eff}\simeq$ 4\parcs5.
To ensure a reasonable $I(m)$ profile, we also impose weak constraints at 
large and small radii, resulting in 50 data points over the range $m=$ 
0\parcs0-29\parcs4.  Falco {\it et al}. (1997) measured the central 
dispersion of G1 to be $\hat{\sigma}_{\rm p} = 279 \pm 12$ km s$^{-1}$
inside a 0\parcs6 ($1.9 h^{-1}$ kpc) radius.  Their data suggest a rise in 
the dispersion at the center, with $\hat{\sigma}_{\rm p} = 316 \pm 14$ 
km s$^{-1}$ inside 0\parcs2, and $\hat{\sigma}_{\rm p} = 266 \pm 12$ 
km s$^{-1}$ outside 0\parcs2.  
Since such a rise is inconsistent with the radial dispersion profiles of 
nearby galaxies and with the subsequent dispersion measurements of
Tonry \& Franx (1998),
we adopt the total binned measurement of 279 km s$^{-1}$.

Grogin \& Narayan (1996) modeled the lensing properties of Q0957+561, using 
two different parametric mass models for G1.  We will compare our results 
with their softened power-law sphere (SPLS) model results, where their 
density profile is $\rho(r)= \rho_0 (1+r^2/r^2_{\rm c})^{-\alpha/2}$.
Their best-fit parameters are ($\alpha=1.92^{+0.08}_{-0.09}$, 
$r_{\rm c}=0\parcs058^{+0{\tiny \farcs053}}_{-0{\tiny \farcs058}}$, 
$\alpha_{\rm E}=2\parcs40^{+0{\tiny \farcs28}}_{-0{\tiny \farcs34}}$,
at 2 $\sigma$), where the deflection parameter $\alpha_{\rm E}$ is related 
to the central density $\rho_0$.  This best-fit model is a poor fit to the 
data ($\chi^2$ per degree of freedom of 6.9), and uses a position for G1 
that has been shown to be incorrect (Bernstein {\it et al.} 1997).  But for
want of a reanalysis of the lensing constraints, we will adopt this 
solution in our models.

Impey {\it et al}. (1998) found that the galaxy PG 1115+080 G is nearly 
circular, and fit well by a de Vaucouleurs profile with 
$R_{\rm eff}=0\parcs59\pm0\parcs06$.  We approximate the surface brightness
of this galaxy by a Hernquist (1990) model with a break radius of 
$a=0\parcs325$ ($R_{\rm eff} \simeq 0\parcs59$), modeled as a profile 
$I(m)$ with 21 bins over the range $m=$ 0\parcs0-2\parcs1.  Tonry (1998) 
measured the central dispersion of G to be 
$\hat{\sigma}_{\rm p} = 281\pm 25$ km s$^{-1}$ inside a 0\parcs6 
($1.7 h^{-1}$ kpc) radius.  Impey {\it et al}. (1998) fit the lensing 
constraints with three standard mass models for G: a singular isothermal 
sphere ($\alpha=2$, $r_{\rm c}=0\arcs$), a modified Hubble profile 
($\alpha=3$, $r_{\rm c}=0\parcs2$), and a constant mass-to-light ratio 
model.  For all the models, the total projected mass inside $R=1\parcs15$ 
was found to be (1.24-1.39)$ \times 10^{11} h^{-1} M_{\odot}$, depending on
the mass model assumed for a nearby galaxy group.

\subsection{Self-Similarity of Kinematic Profiles}

Bender, Saglia, \& Gerhard (1994) measured the LOSVD out to 
$\sim R_{\rm eff}$ for a large, ``unbiased'' sample of galaxies, and
derived the profiles $\hat{v}_{\rm p}(R)$, $\hat{\sigma}_{\rm p}(R)$,
$h_3(R)$, and $h_4(R)$.  For our data set, we take from this sample 80 
profiles from 28 elliptical galaxies.  We find the rms projected velocity 
$\tilde{v}_{\rm rms}$ by numerically integrating the positive line profile 
$\tilde{\cal L}(v_{\rm p}$) with fit parameters \{$\hat{v}_{\rm p}$, 
$\hat{\sigma}_{\rm p}$, $h_3$, $h_4$\} (see van der Marel \& Franx 1993 
\S 2.4).  To make a scale-free comparison of the profiles, we renormalize
to the central $\tilde{v}_{\rm rms}$ (inside $0.14 R_{\rm eff}$, for direct
comparison with the central $\hat{\sigma}_{\rm p}$ measurement of Q0957+561 
G1; or inside $1.1 R_{\rm eff}$ for PG 1115+080 G), and rescale radially by
$R_{\rm eff}$.  We use both major and minor axis data, mapping them to the
intermediate radius $m$.  We combine the data in radial bins, spaced such 
that the number of points in each bin is nearly constant ($\simeq 36$).  To
improve the large-radius constraints, we divide the last bin (0.7-1.8 
$R_{\rm eff}$) into three bins, with 12-13 points in each bin.  As shown in
Figure 1b for the case normalized to Q0957+561 G1, the resulting (rescaled) 
``mean $v_{\rm rms}$ profile'' is nearly constant 
($\tilde{v}_{\rm rms} \simeq 0.9 \pm 0.1$ at $1.5 R_{\rm eff}$), while the 
dispersion $\tilde{\sigma}_{\rm p} \equiv \langle (v_{\rm p}-\tilde{v}_{\rm p})^2\rangle^{1/2}$
decreases with radius ($\tilde{\sigma}_{\rm p} \simeq 0.6 \pm 0.1$ at 
$1.5 R_{\rm eff}$) due to increasing rotational support in the outer parts 
of the low-luminosity galaxies (see Davies {\it et al}. 1983; Fisher, 
Illingworth, \& Franx 1995).  We thus find a remarkably universal, flat rms
velocity profile inside $1.5 R_{\rm eff}$, for all elliptical galaxies 
regardless of other properties.  Note that $v_{\rm rms}$ is not only less 
variable than $\hat{\sigma}_{\rm p}$, but is also a better physical probe 
of the gravitational potential.  The corrections in the mean profiles due 
to $h_3$ and $h_4$ are small, so that 
$\tilde{\sigma}_{\rm p} \simeq \hat{\sigma}_{\rm p}$ to $\sim 0.3\%$ 
accuracy, and $\tilde{v}_{\rm rms} \simeq (\hat{v}_{\rm p}^2 + \hat{\sigma}_{\rm p}^2)^{1/2}$
to $\sim 5\%$.

To produce the mean $h_4$ profile, we use the same binning procedure as for
$v_{\rm rms}$ (see Figure 1a).  We also add data from a sample of six 
galaxies (Carollo {\it et al}. 1995; Statler, Smecker-Hane, \& Cecil 1996; 
Gerhard {\it et al}. 1998) for which $h_4$ has been measured at larger 
radii (1.7-$4.9 R_{\rm eff}$); we combine these points into three 
additional bins, with $\simeq 12$ points in each bin.  Note that these 
points are not a large sample, and comparisons of different authors' data 
sets suggest that the errors in large radius velocity data are generally 
underestimated.  The mean and dispersion range from 
$h_4 \simeq 0.002\pm0.03$ in the central bin to $0.02\pm0.06$ at 
$1.5 R_{\rm eff}$, and the LOSVD is everywhere consistent with Gaussianity 
($h_4=0$).

Because our mean profiles are derived from a general data set of elliptical
galaxies, they may not accurately represent a brightest cluster galaxy (BCG)
like Q0957+561 G1.  To gauge the magnitude of any systematic errors thereby 
caused, we repeat our procedure with 42 velocity profiles from a set of 12 
BCGs from Fisher {\it et al}. (1995), who find that the radial slopes of 
$\hat{\sigma}_{\rm p}$ are similar to those of a sample of normal 
elliptical galaxies.  Since their data do not include measurements of $h_3$
and $h_4$, we use the approximations mentioned above for 
$\tilde{v}_{\rm rms}$ and $\tilde{\sigma}_{\rm p}$.  As shown in Figure 1c,
the resulting BCG $\hat{\sigma}_{\rm p}$ and $v_{\rm rms}$ profiles are 
consistent to $\sim$ 5-10\% accuracy with those from the Bender {\it et al}.
(1994) data set, although small-number statistics make the significance of 
the differences difficult to interpret.  We also examine the Bender {\it et
al}. (1994) data for any systematic correlations with other galaxy 
properties (the absolute magnitude $M_B$, the dimensionless rotation 
$[\hat{v}_{\rm m}/\langle\hat{\sigma}_{\rm p}\rangle]^*$, and the stellar 
projected axis ratio $q_*$).  While we find indications that ``BCG-like'' 
galaxies have systematically lower $h_4$ and higher 
$\tilde{\sigma}_{\rm p}$ at large radii, they show no clear difference for 
$\tilde{v}_{\rm rms}$.  Although there are insufficient data available to 
construct BCG mean profiles, we can still use them to {\it estimate} the 
systematic corrections to our modeling results for Q0957+561 G1 (see \S 4).

\longpagen[8.41in]
\newpage
\normalpage
{\plotfiddle{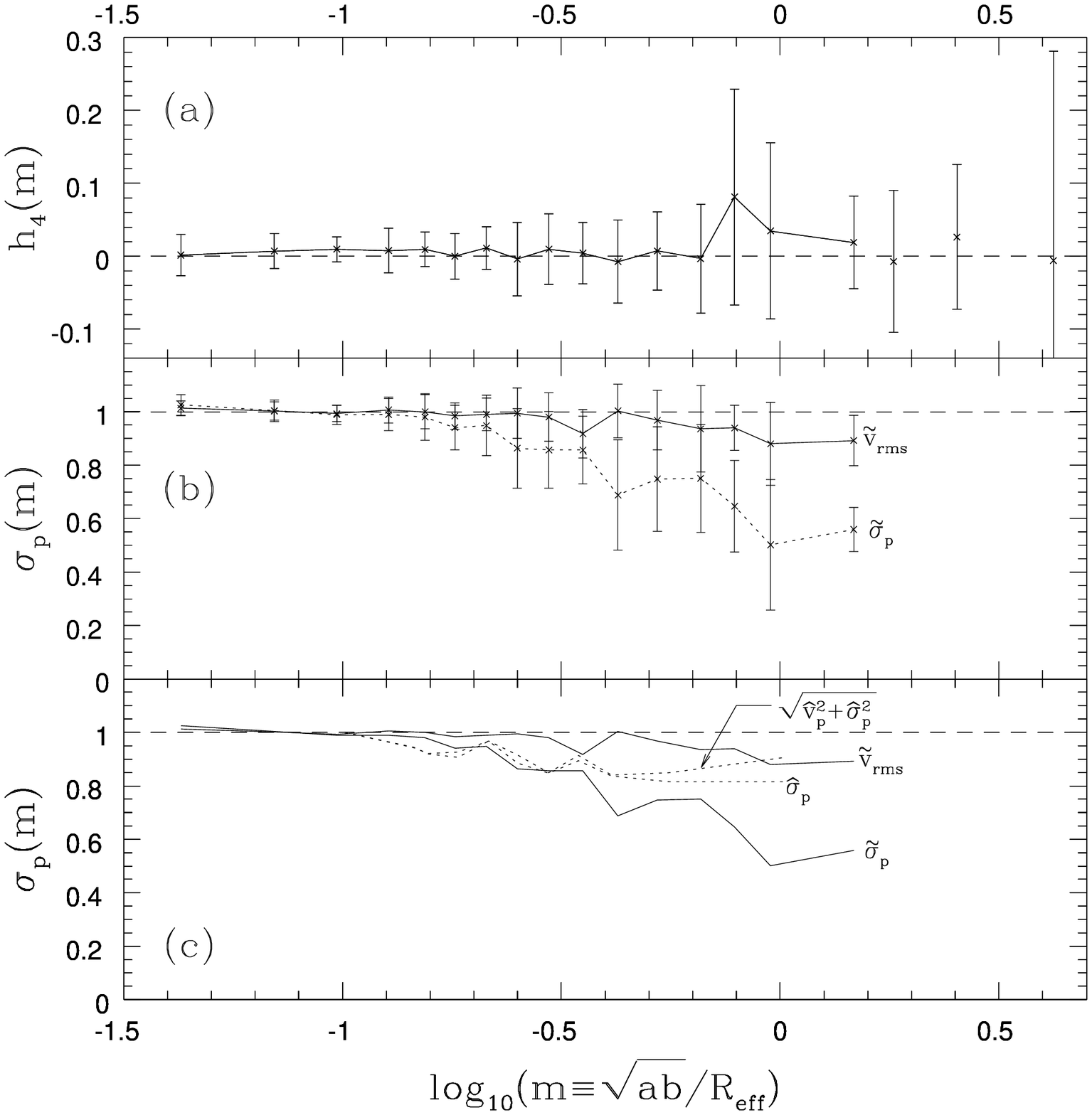}{6.78in}{0}{90}{90}{-45}{-137}}

{\center \small {\bf F{\scriptsize \bf IG}. 1.---}
Mean profiles computed from elliptical galaxy data.  Error bars show the
dispersion about the mean.  (a) Fourth-order Gauss-Hermite moment.  A solid 
line connects the Bender {\it et al}. (1994) data.  (b) Velocity profiles,
from the Bender {\it et al}. (1994) data, normalized to the total velocity 
dispersion inside the radius $m=0.14$.  Shown are both the rms velocity 
({\it solid line}) and the velocity dispersion ({\it dotted line}).
(c) Velocity profiles, from the Bender {\it et al}. (1994) ({\it solid 
lines}) and the BCG ({\it dotted lines}) data.  For clarity, the scatter is 
not shown.
}

\newpage
\longpagen[8.43in]
\section{METHODS}

Schwarzschild (1979) described, and Richstone \& Tremaine (1984) extended,
a completely general method of dynamical modeling, where a galaxy is built 
from a library of representative orbits, each weighted with an occupancy 
number.  The weights are adjusted so that the model fits a set of 
observational constraints --- typically the surface brightness and 
line-of-sight velocities of a galaxy.  By construction, the method arrives
at a solution that is a physical system of non-negative orbits (thereby 
avoiding the problems with using the Jeans equations).  Unlike other common
modeling methods, the method requires explicit knowledge of neither the 
integrals of motion nor the form of the distribution function.  Recent 
efforts at galaxy modeling have employed sophisticated variants of the 
method that include fits to higher-order velocity moments ({\it e.g.}, Rix 
{\it et al}. 1997; van der Marel {\it et al}. 1998).

As discussed in \S 2.1, we adopt the SPLS family of density profiles for 
our models, along with the Hernquist (1990) mass model.  The initial radii 
$r_{0k}$ of the orbits are logarithmically spaced in $r_0$, and the energy 
$E_k$ of each orbit is selected to correspond to that of a circular orbit
at this radius, $\Phi(r_0) + v_{\rm c}^2(r_0)/2$.  For a singular 
isothermal potential, the spacing is uniform in energy.  The angular 
momentum $L_k$ of the orbit is selected randomly from the range 
$[0,L_{\rm max}]$, where $L_{\rm max} = r_0 v_{\rm c}(r_0)$.  This 
procedure ensures dense, uniform coverage of the $(E,L)$ phase space.
The model observables ${\bf y}^{\rm m}$ are given by the orbit weights
$\bf{w}$ and the orbit projection ``kernels'' $\bf{K}^{\rm m}$, which are 
averaged over time and over all spherical-polar viewing angles 
($\theta,\phi$):
$y^{\rm m}_i = \sum_k w_k^2 \langle K_{ik}^{\rm m}\rangle_{t,\theta,\phi}$.
For example, the kernel for the angle-averaged surface density of 
an orbit at radius $r$ is given by
$\langle K^{I(R)}\rangle_{\theta,\phi} = (2 \pi r \sqrt{r^2-R^2})^{-1}$.
The orbit is then run forward in time for one radial period $T_r$, 
and the final kernel is found by averaging over time:
$\langle K \rangle_{t,\theta,\phi} = T_r^{-1} \int_0^{T_r} \langle K\rangle_{\theta,\phi}(t) dt$.
To calculate the Gauss-Hermite velocity moments, we calculate the LOSVD 
${\cal L}(v_{\rm p})$ in 41 velocity bins from $v_{\rm p} = 0$ to 
$v_{\rm max}$, where the maximum velocity $v_{\rm max}$ is given by the 
largest velocity attainable on a radial orbit by the highest-energy orbit.
We then perform a nonlinear least-squares fit to find $\hat{\gamma}$ and 
$\hat{\sigma}_{\rm p}$, and then use equation (1) to find $h_l$.

Since the model is typically underconstrained, we fit the model observables 
$\bf{y}^{\rm m}$ to the data $\bf{y}^{\rm d}$ using the statistic
$\chi^2 = \sum_i (y^{\rm m}_i-y^{\rm d}_i)^2/\sigma_i^2$, while optimizing 
a smoothing function, the entropy $S = \sum_k w_k^2 \ln w_k^2$.  This 
corresponds to minimizing the function $f \equiv \chi^2+\lambda S$, where
the Lagrangian multiplier $\lambda$ is a smoothing factor.  During our 
modeling runs, we reduce $\lambda$ slowly from 1 to $10^{-5}$ to arrive at 
the limiting case where no smoothing is imposed.  Since the velocities 
scale linearly with the total mass of the galaxy, we can leave the mass 
dispersion parameter ($\sigma_0^2 \equiv 2 \pi G \rho_0 r_{\rm c}^2$ for 
the SPLS models) free to vary in the fit.  We can also enforce isotropy by 
minimizing the function
$f_i \equiv \sum_i [(\nu v_r^2)_i - (\nu v_t^2)_i]^2/[(\nu v_r^2)_i + (\nu v_t^2)_i]^2$.
We use the conjugate gradient method (Press {\it et al}. 1992), with first 
and second derivative information, to minimize $f$.

We test our methods on a self-consistent (constant mass-to-light ratio) 
isotropic Hernquist (1990) galaxy, with mass density profile 
$\rho(r) = M_0 a (2 \pi r)^{-1} (r+a)^{-3}$.  We fit the exact analytic
stellar surface brightness profile $I(R)$ and the projected stellar 
velocity dispersion profile $\sigma_{\rm p}(R)$.  Both profiles are 
measured in 21 annuli from $R = 0$ to $R = 16 a$, and are assigned 10\% 
measurement errors.  There are 2000 orbits, spaced with initial radii from
$0.07 a$ to $221 a$ ($R_{\rm eff} \simeq 1.8 a$), resulting in a radial 
coverage from 0 to $442 a$.  There are a variety of solutions consistent 
with the data, including the self-consistent isotropic solution.  Although 
the more anisotropic solutions generally show more pronounced deviations 
from Gaussianity in their LOSVDs, isotropy does not necessarily imply 
$h_4=0$, and vice-versa (see Figure 2).

{\plotfiddle{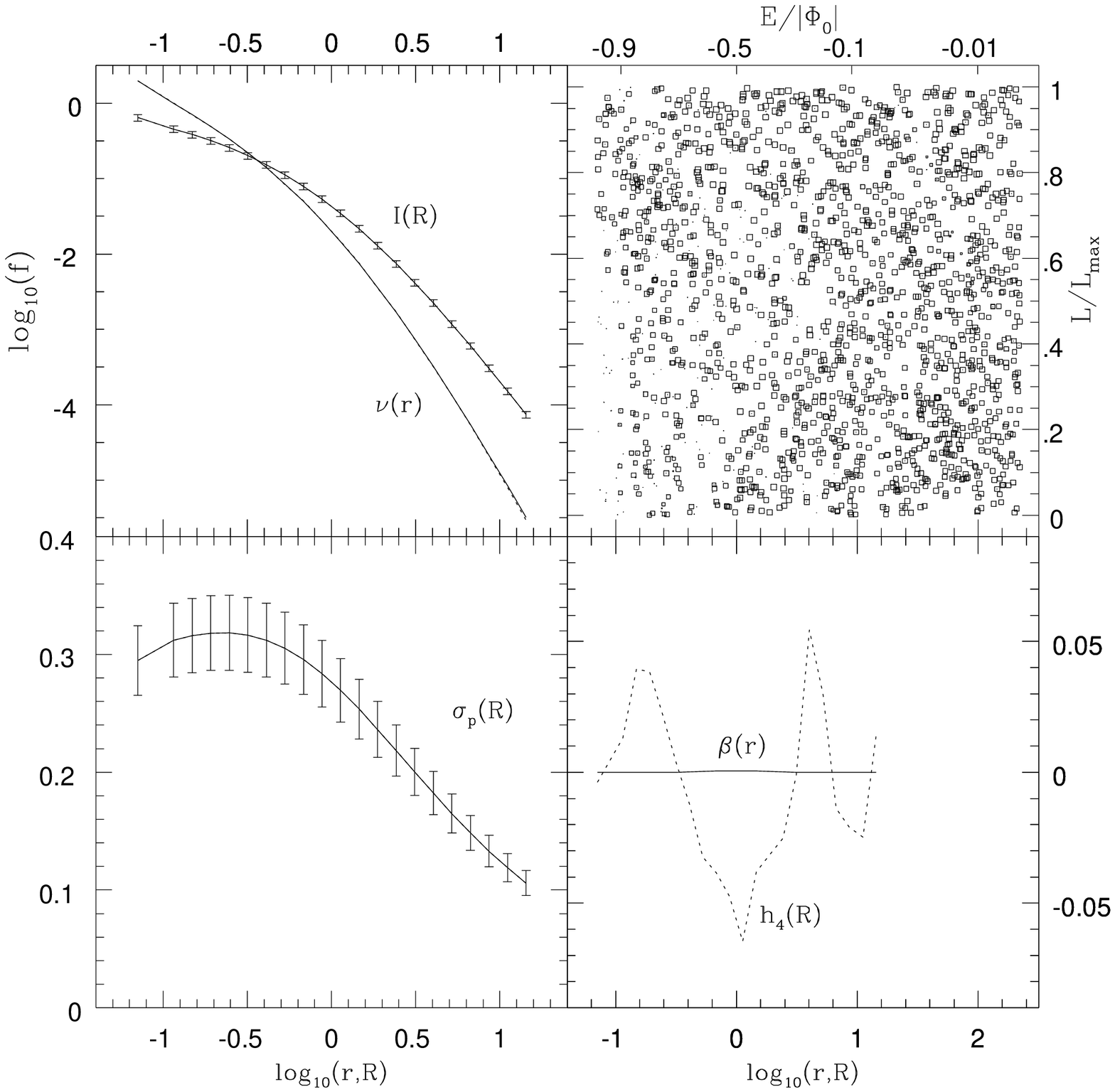}{5.25in}{0}{67}{67}{15}{-112}}

{\center \small {\bf F{\scriptsize \bf IG}. 2.---}
Isotropic model solution to simulated Hernquist (1990) galaxy data.
{\it Upper left:} Model ({\it dotted lines}) and data ({\it solid lines}),
for surface brightness (with error bars) and luminosity density.  
{\it Lower left:} Same, for projected stellar velocity dispersion.
{\it Upper right:} Orbit weights, in energy-angular momentum phase space.
The area of each square is proportional to the logarithm of the orbit's 
weight.  The bottom axis shows the orbit's initial radius, and the top axis 
its energy.  {\it Lower right:} Anisotropy parameter ({\it solid line}) and 
fourth-order Gauss-Hermite velocity moment ({\it dotted line}), as a 
function of radius. The break radius is $a = 1$ ($R_{\rm eff} \simeq 1.8$).
}

\normalpage
\newpage
\longpagen[8.97in]
\section{Q0957+561 RESULTS}

We next model the galaxy G1 in Q0957+561, fitting only the measured data: 
$I(R)$ and central $\hat{\sigma}_{\rm p}$ (see \S 2.1).  We use the 
best-fit SPLS mass model from GN, with ($\alpha=1.92$, 
$r_{\rm c}=0\parcs058$).  There are 2000 orbits with initial radii from 
0\parcs07 to $221\arcs$, resulting in radial coverage from $0\arcs$ to 
$361\arcs$.  A wide range of solutions fits the data exactly, and we find 
1-$\sigma$ limits on the mass dispersion parameter of 
$\sigma_0 = 295^{+143}_{-121}$ km s$^{-1}$, defined by the 
$\Delta \chi^2 = 1$ boundary about the minimum $\chi^2$ (see Figure 3).
Such a large range of possible solutions corresponds to a 73\% uncertainty 
in the mass of G1, but the extreme solutions show radical departures from a
constant velocity dispersion profile, and from nearly-Gaussian LOSVDs (see 
Figure 4). For example, a very massive solution has nearly circular orbits
at large radii, so that there are few plunging radial orbits to produce 
large velocities at the galactic center (see Figure 6).  This behavior shows
up as a velocity dispersion profile that rises with radius, and has a 
negative $h_4$ moment (flat-topped LOSVD) at large radii (see Figure 5).

To reject such solutions, we impose our ``mean profile'' constraints on 
$\hat{\sigma}_{\rm p}(R)$ and $h_4(R)$ (see \S 2.2; note that the spherical 
symmetry in our model implies $v_{\rm rms} \simeq \hat{\sigma}_{\rm p}$).
We find that the range of viable solutions is dramatically reduced (Fig. 3), 
with new 1-$\sigma$ limits on $\sigma_0$ of $280^{+19}_{-26}$ km s$^{-1}$,
and 2-$\sigma$ limits of $280^{+29}_{-34}$ km s$^{-1}$.  
Some of these solutions may 
not be dynamically stable, but incorporating stability criteria into our 
model fitting is beyond the scope of this project.

{\plotfiddle{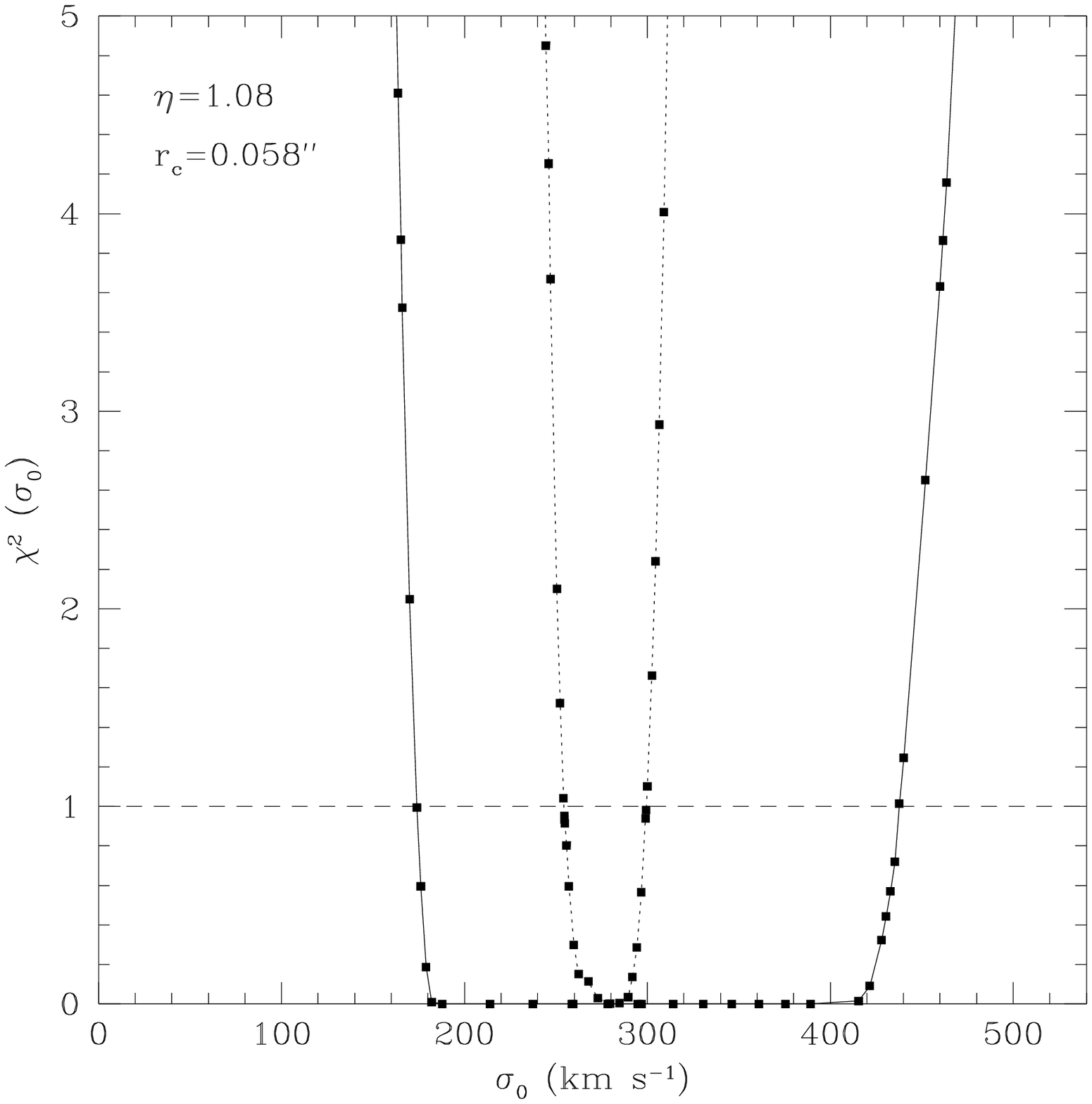}{3.48in}{0}{50}{50}{76}{-94}}

{\center \small {\bf F{\scriptsize \bf IG}. 3.---}
Chi-square of the best-fit lens mass model to the Q0957+561 G1 data (surface 
brightness and central projected velocity dispersion), as a function of the 
mass dispersion ({\it solid line}).  Also shown are the fits including the 
mean $\hat{\sigma}_{\rm p}(R)$ and $h_4(R)$ constraints ({\it dotted 
line}).
}

Since the reduced range of the solutions depends on the application of our 
mean profile constraints, we next investigate the sensitivity of the 
results to our main concerns about these constraints.  First, we remove the 
constraints on $h_4$ at large radii ($R > 1.8 R_{\rm eff}$), where the 
reliability of the local galaxy data is questionable.  This increases the 
best-fit $\sigma_0$ and its uncertainty
by less than 1 km s$^{-1}$.  Second, we estimate 
the effect of a systematic bias.  Since we found in \S 2.2 that BCGs may 
have systematically lower $h_4$ at large radii, we examine an extreme case 
where we set $h_4 = -0.06$ for $R>1\parcs2$.  We find new 1-$\sigma$ limits
of $\sigma_0= 287^{+15}_{-18}$ km s$^{-1}$, indicating that the systematic 
correction for galaxy type would increase $\sigma_0^2$ (and $H_0$) by at 
most 5\%.  In summary, we find 1-$\sigma$ limits on $\sigma_0$ of 
$280^{+19}_{-26}$ km s$^{-1}$, which corresponds to a 16\% uncertainty in
the mass of G1.
Part of the uncertainty in $\sigma_0$ is due to the 
measurement error of the central $\hat{\sigma}_{\rm p}$ (12 km s$^{-1}$),
while part is due to the systematic uncertainty in converting from 
$\hat{\sigma}_{\rm p}$ to $\sigma_0$ ($\sim 19$ km s$^{-1}$).  This contrasts 
strongly with the conversion from GN, which implies (by assuming
the anisotropy $\beta(r)$ to be constant and near zero) 1-$\sigma$ limits of
$\sigma_0 = 290^{+12}_{-13}$ km s$^{-1}$.  Their reported systematic 
uncertainty of 2 km s $^{-1}$ is clearly underestimated, given that our 
best dynamical model is systematically different from theirs by a factor 
five times larger.

{\plotfiddle{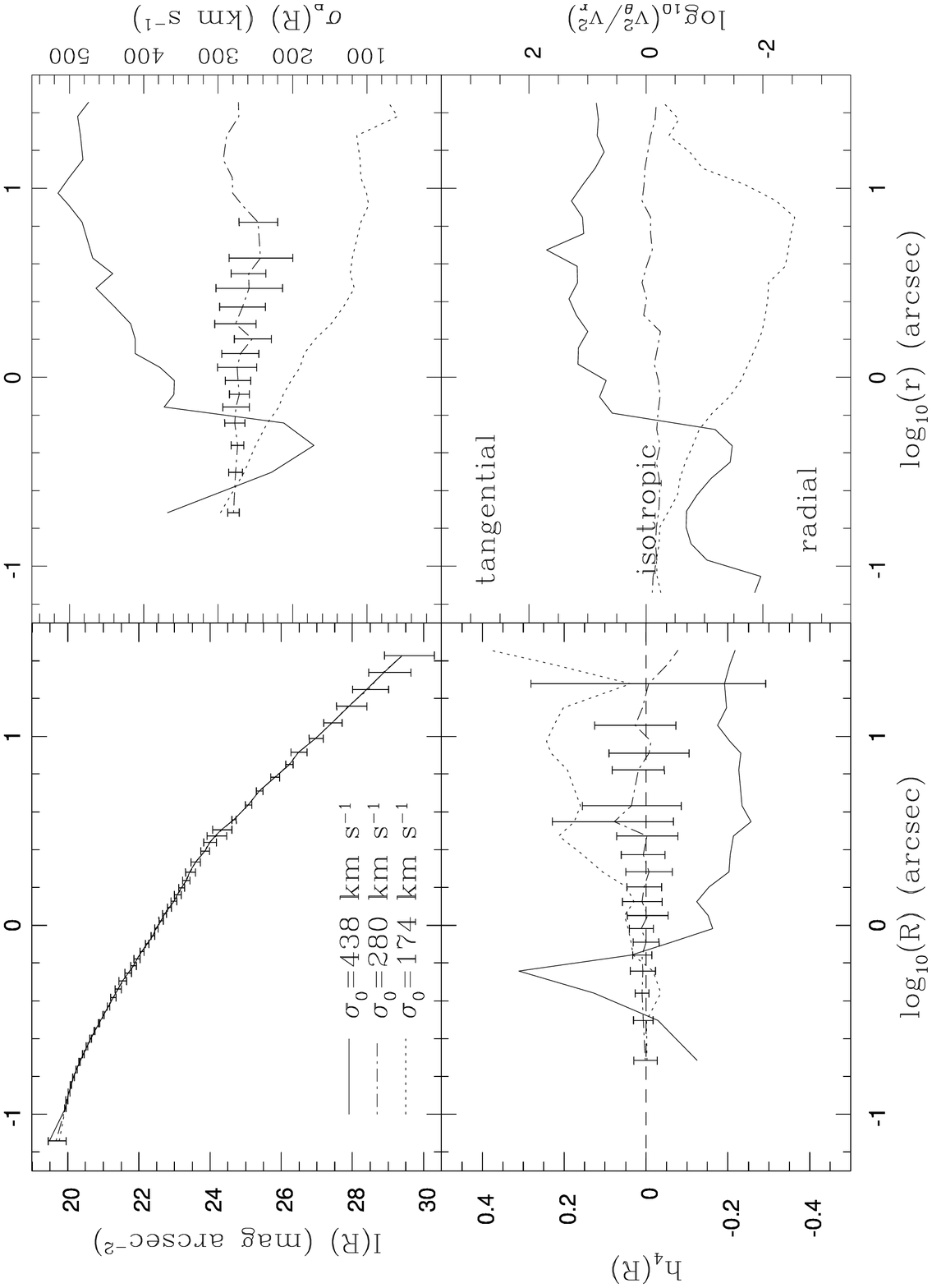}{4.35in}{-90}{60}{60}{-12}{338}}

{\center \small {\bf F{\scriptsize \bf IG}. 4.---}
Solutions for Q0957+561 G1, for several mass dispersions.  The solutions are 
all acceptable ($\Delta \chi^2 < 1$), given only the surface brightness and 
central velocity dispersion data.  {\it Upper left:} Surface brightness 
profile, where the error bars show data from Bernstein {\it et al}. (1993, 
1997).  {\it Lower left:} Fourth-order Gauss-Hermite moment profile, with 
the ``mean profile'' shown by error bars (see \S 2.2).  {\it Upper right:} 
Velocity dispersion profile, with the mean profile shown by error bars.
{\it Lower right:} Anisotropy profile.
}

\longpagen[8.43in]
\newpage
\normalpage
{\plotfiddle{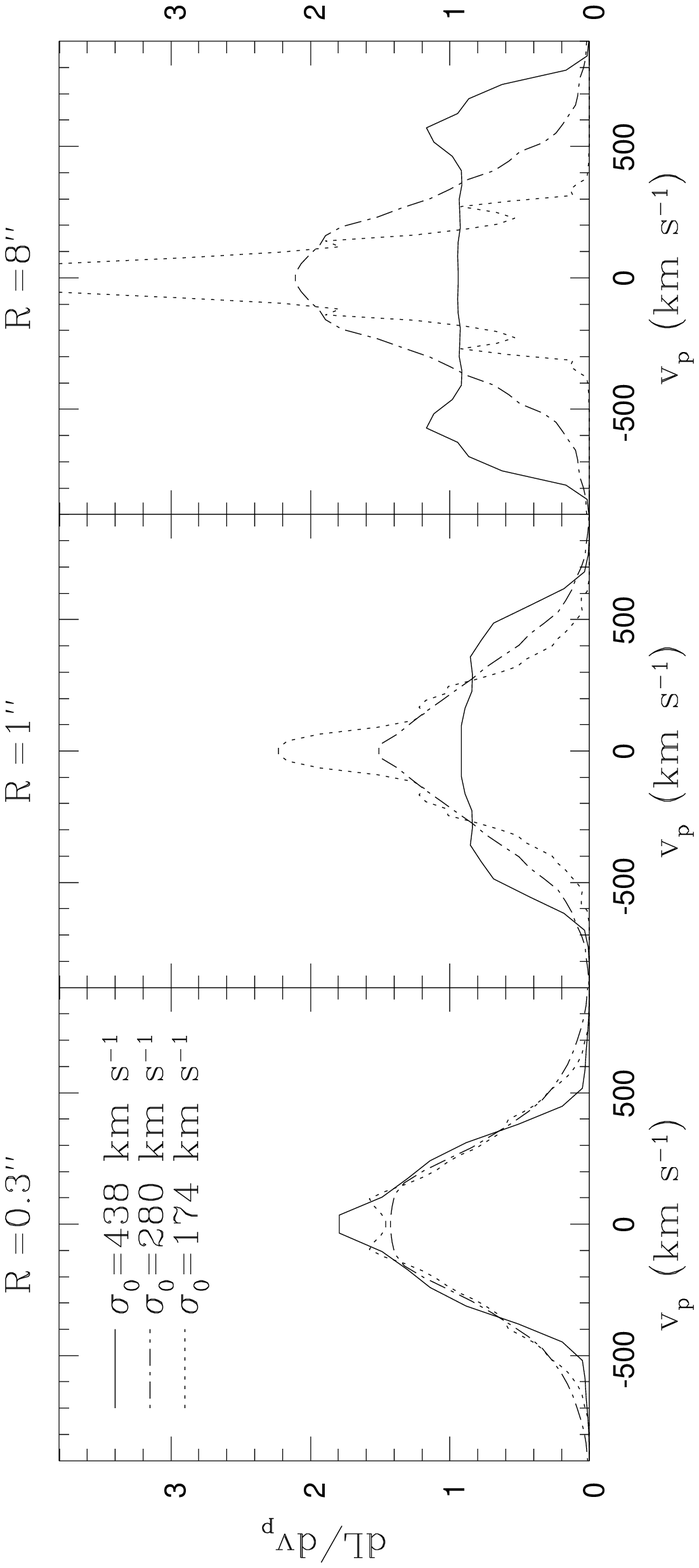}{3.14in}{-90}{75}{75}{-75}{250}}

{\center \small {\bf F{\scriptsize \bf IG}. 5.---}
Line-of-sight velocity distributions, for the same solutions, at several 
radii.
}

{\plotfiddle{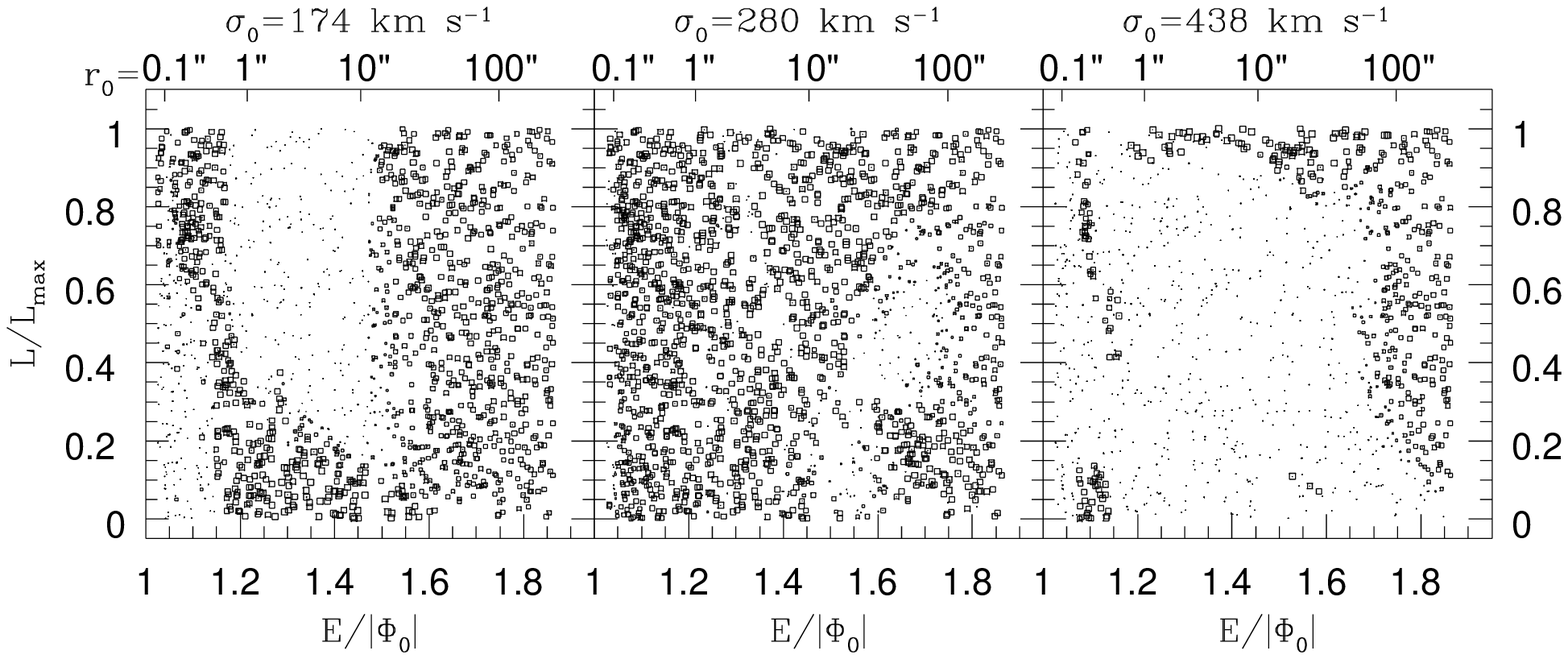}{3.55in}{0}{100}{100}{-83}{-490}}
{\center \small {\bf F{\scriptsize \bf IG}. 6.---}
Orbit weights for the same solutions, plotted in energy-angular momentum 
phase space.  The area of each square is proportional to the logarithm of 
the orbit's weight.  The top axis shows the initial radius.
}

\newpage

We next compare our results with the Q0957+561 lens models of GN to produce 
new bounds on the Hubble constant $H_0$.  Given their best fit mass model 
$(\alpha=1.92, r_{\rm c}=0\parcs058)$, GN found 1-$\sigma$ limits on the 
deflection parameter $\alpha_{\rm E}$ of $2\parcs40\pm0\parcs07$.  As 
mentioned in \S 1, the conversion of $\alpha_{\rm E}$ to the G1 mass 
dispersion $\sigma_0$ is subject to a well-known degeneracy between the
galaxy mass and the cluster mass: one can add a cluster with a convergence 
$\kappa$, and decrease the galaxy mass by the factor $(1-\kappa)$.  Since 
this degeneracy affects none of the image observables but the time delay, 
$H_0$ is systematically uncertain by the same factor $(1-\kappa)$.  Thus, 
with the time delay measurement of Kundi\'{c} {\it et al.} (1997), and 
density parameter $\Omega_0=1$, the GN results imply 
$\sigma_0 = (324\pm4)\sqrt{1-\kappa}$ km s$^{-1}$ and 
$H_0 = (82\pm2) (1-\kappa)$ km s$^{-1}$ Mpc$^{-1}$.
Our dynamical constraints on $\sigma_0$ therefore put 1-$\sigma$ constraints on
$\kappa$ of $0.25^{+0.14}_{-0.10}$, implying $H_0 = 61^{+9}_{-11}$ 
km s$^{-1}$ Mpc$^{-1}$; and 2-$\sigma$ constraints of 
$\kappa = 0.25^{+0.19}_{-0.16}$, implying $H_0 = 61^{+13}_{-15}$ km s$^{-1}$ 
Mpc$^{-1}$ (see Figure 7).
Note that Tonry \& Franx (1998) measured a central velocity dispersion 
for G1 of $\hat{\sigma}_{\rm p} = 288 \pm 9$ km s$^{-1}$;
although their aperture is different from that used in our dynamical models,
their measurement would roughly imply $\kappa \simeq 0.20\pm0.12$ and
$H_0 \simeq 65 \pm 10$ km s$^{-1}$ Mpc$^{-1}$.

For comparison, we examine complementary studies of this system.  Fischer 
{\it et al}. (1997) used the weak lensing of background galaxies to 
determine the surface density $\Sigma(R)$ of the cluster.  Using their 
parameterized model fit to $\Sigma$ and their stated uncertainties, we find
1-$\sigma$ constraints on the convergence $\kappa$ of $0.19^{+0.33}_{-0.09}$,
and 2-$\sigma$ bounds of (0.06-1.00).  
Similarly, Kundi\'{c} {\it et al}. (1997) estimated 
$\kappa=0.22\pm0.14$ (2 $\sigma$) from this data, but 
they neglected the uncertainties in the cluster position.  Note that these 
estimates of $\kappa$ were derived with the cluster centered on the galaxy 
G1, rather than at the real mass centroid of the cluster, and also that the 
proximity of the cluster to G1 invalidates the description of the potential 
as simply a convergence $\kappa$ and a shear $\gamma$ (see Kochanek 1991).
Kundi\'{c} {\it et al}. (1997) also warn that an error in the assumed mean 
redshift of the background galaxies can affect the derived $H_0$ 
significantly.  With these {\it caveats} in mind, we find that the 1-$\sigma$
values for $\kappa$ from the Fischer {\it et al}. (1997) models imply 
$\sigma_0= 291^{+16}_{-59}$ km s$^{-1}$ for the best-fit GN model, and 
$H_0=66^{+8}_{-27}$ km s$^{-1}$ Mpc$^{-1}$ (see Fig. 7).  In a complementary
study, Chartas {\it et al}. (1998) determined the mass of the cluster from 
its gaseous X-ray emission.  Their results imply $\kappa = 0.11 \pm 0.04$,
$\sigma_0 = 306 \pm 17$ km s$^{-1}$, and $H_0 = 73\pm6$ km s$^{-1}$ 
Mpc$^{-1}$ (2 $\sigma$), but they include no estimate of their systematic 
uncertainties.

Both of these direct constraints on $\kappa$ are consistent with the
constraints implied by our dynamical models of G1,
and could in principle be combined with them to further constrain $H_0$,
though they are currently too uncertain to add any useful information.
So we find that the GN lens model, combined with the dispersion measurement of G1,
permits a determination of $H_0$ to 16\% accuracy.
Although better lens models may significantly shift the best-fit $H_0$ value,
they will have little effect on the
uncertainties, which are dominated by the uncertainty in $\kappa$; we have 
already assumed a perfect determination of the parameters ($\alpha$, 
$r_{\rm c}$) from the GN models, and a perfect determination of 
$\alpha_{\rm E}$ would tighten the 1-$\sigma$ limits on $H_0$ by only
$\sim 0.1$ km s$^{-1}$ Mpc$^{-1}$.  
To find $H_0$ to significantly better 
accuracy from this system, we would need much better direct constraints on the 
cluster mass distribution, and/or highly accurate velocity data from the 
galaxy at larger radii.

\newpage
{\plotfiddle{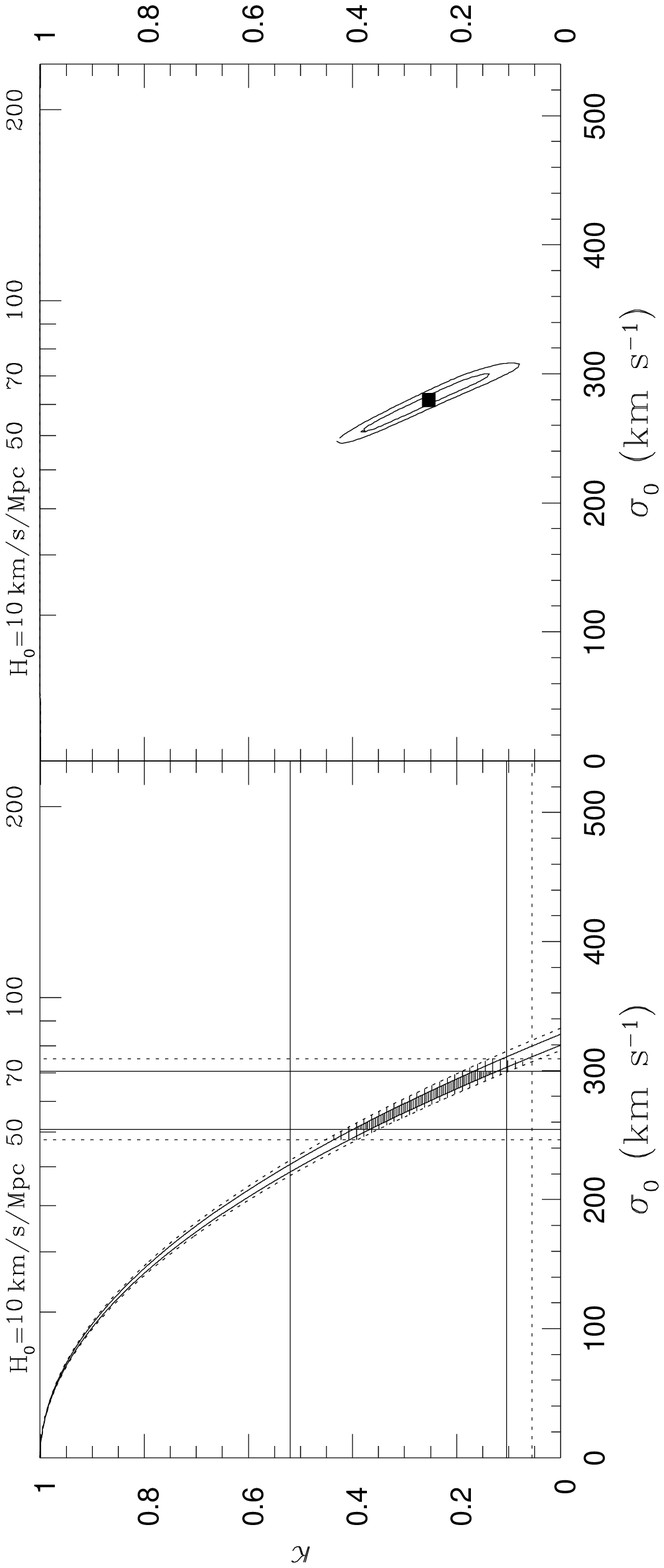}{2.77in}{-90}{70}{70}{-50}{221}}
{\center \small {\bf F{\scriptsize \bf IG}. 7.---}
Constraints on galaxy mass dispersion $\sigma_0$, cluster convergence 
$\kappa$, and Hubble constant $H_0$.  The left panel shows a diagram of the
independent constraints.  The solid lines show 1-$\sigma$ bounds on 
$\sigma_0$ and $\kappa$, and the dotted lines show 2-$\sigma$ bounds.  The 
curved lines show the constraints on $\sigma_0$ and $\kappa$ from the GN 
lens models.  The horizontal lines show the constraints on $\kappa$ from 
Fischer {\it et al}. (1997).  The vertical lines show our constraints on 
$\sigma_0$.  The approximate region of parameter space permitted at the 1-$\sigma$ 
level is indicated by dark shading; the 2-$\sigma$ region by light shading.
The right panel shows a contour plot of the permitted region
(including the GN and $\sigma_0$ constraints), 
with 1-$\sigma$ and 2-$\sigma$ bounds.  The point marks the best-fit solution.
}

\section{PG1115+080 RESULTS}

We next model the galaxy G in PG 1115+080 to determine if the different 
lens mass profiles explored by Impey {\it et al}. (1998) are consistent 
with the measurements by Impey {\it et al}. (1998) of $I(R)$ and by Tonry 
(1998) of the central $\hat{\sigma}_{\rm p}$ (see \S 2.1).  If any mass 
model can be ruled out, so can its corresponding value of $H_0$.  As in the 
case of Q0957+561, a single central $\hat{\sigma}_{\rm p}$ measurement will 
give us little information about the galaxy mass profile, so we again 
impose mean profile constraints on $v_{\rm rms}(R)$ and $h_4(R)$, 
normalized to this galaxy (see \S 2.2).  

We fit a singular isothermal mass 
model to these data,
and although we have not exhaustively explored the entire range of possible solutions,
we estimate the projected mass inside $R=1\parcs15$ to be
$M=1.7^{+0.3}_{-0.7}\times 10^{11} h^{-1} M_\odot$,
which is consistent with the lens model's implied 
$M=(1.25\pm0.02)\times 10^{11} h^{-1} M_\odot$
(for the case where the nearby galaxy group is modeled as a singular 
isothermal sphere).  Similarly, we fit the modified Hubble model and find 
$M = (1.6\pm0.5)\times 10^{11} h^{-1} M_\odot$, which is 
consistent with the lens model fit mass.
To test the constant mass-to-light ratio hypothesis, we fit a 
Hernquist (1990) model with a 
break radius of $a=0\parcs325$ (to match the measured effective radius of 
$R_{\rm eff}=0\parcs59$), and find
$M=1.3^{+0.3}_{-0.4} \times 10^{11} h^{-1} M_\odot$, which is
also consistent with the lens model fit mass.
Our dynamical models are also consistent with the lens model results
when the group is modeled as a point mass,
in which case $M\simeq 1.4\times 10^{11} h^{-1} M_\odot$.
To compare the relative likelihood of the three mass models, 
we impose the lens model mass normalization on each of them, and find that
our dynamical solutions are statistically indistinguishable ($\Delta \chi^2 < 1$).

With the measured time delay (Schechter {\it et al}. 1997 ; Barkana 1997b),
for $\Omega_0=1$ and a singular isothermal group model, 
the singular 
isothermal galaxy model gives $H_0=44\pm4$ km s$^{-1}$ Mpc$^{-1}$; the 
modified Hubble profile model, $61\pm5$ km s$^{-1}$ Mpc$^{-1}$; and the 
constant $M/L$ model, $65\pm5$ km s$^{-1}$ Mpc$^{-1}$ (Impey {\it et al}. 
1998).  
Treating the group as a point mass 
increases $H_0$ by $\sim 10\%$.  
Since our stellar dynamical models do not rule out any of the lens models,
a large range of values for $H_0$ is still permitted by the system.
As in the case of Q0957+561, stronger constraints on the mass 
distribution of the system 
(e.g., from LOSVD measurements of the lens galaxy, or from
more detailed observations of the Einstein ring)
will be necessary to break the model degeneracies.

\section{CONCLUSIONS}

Using very general orbit modeling methods, we have examined the uncertainty 
in the mass of the lens galaxies Q0957+561 G1 and PG 1115+080 G, given 
observations of their central projected stellar velocity dispersions
$\hat{\sigma}_{\rm p}$ (Falco {\it et al.} 1997; Tonry 1998).  As many past 
studies have shown, such a measurement alone is inadequate to strongly 
constrain the galaxy's mass.  In order to put additional realistic 
{\it a priori} constraints on the galaxy's properties, we have derived 
``mean profiles'' of the rms projected velocity $v_{\rm rms}(R)$ and the 
fourth-order Gauss-Hermite moment $h_4(R)$ from a large sample of nearby 
elliptical galaxies.  These mean profiles prove to be remarkably 
self-similar, even over a large range of galaxy types.  This universality 
is not too surprising, given the homology of early-type galaxies implied by 
the existence of the fundamental plane --- an even stronger implication if 
the central kinetic energy is considered instead of the velocity dispersion
({\it e.g.}, Busarello {\it et al}. 1997).

For Q0957+561 G1, given the best-fit SPLS mass model from Grogin \& Narayan 
(1996), with only the surface brightness profile $I(R)$ and central 
$\hat{\sigma}_{\rm p}$ as constraints, we find 1-$\sigma$ limits on the 
mass dispersion $\sigma_0$ of $295^{+143}_{-121}$ km s$^{-1}$.  The addition 
of the mean profile constraints reduces this permitted range to 
$280^{+19}_{-26}$ km s$^{-1}$.  In conjunction with the GN lens model 
constraints, this implies a cluster convergence of 
$\kappa=0.25^{+0.14}_{-0.10}$, 
which is consistent with the constraints on $\kappa$ from other independent
studies.
Using the time delay measurement of Kundi\'{c} {\it et al}. (1997), 
we find 1-$\sigma$ 
limits on $H_0$ of $61^{+9}_{-11}$ km s$^{-1}$ Mpc$^{-1}$, and 2-$\sigma$ 
limits of $61^{+13}_{-15}$ km s$^{-1}$ Mpc$^{-1}$.  {\it Thus, current 
measurements of the lens system Q0957+561 do not constrain $H_0$ to better 
than 15\%.} To obtain useful limits on $H_0$, we will need better 
constraints on the cluster convergence and/or better velocity profile 
measurements for the galaxy G1.  
We will also need substantial revision of the lens model ---
the GN solutions
fit the lens data poorly, 
inaccurately assume a spherical galaxy,
and use an oversimplified Taylor expansion model of the cluster.
Presumably a more accurate lens model will eventually provide a good fit
to the data and a different value for $H_0$
(see Barkana {\it et al}. 1998 for some improved models),
in which case our dynamical model will still be illustrative of the 
systematic uncertainties expected.

We have also examined the lens galaxy PG 1115+080 G, which has a total mass 
that is relatively well-determined by the lensing constraints, but a radial 
mass distribution that is unconstrained (Impey {\it et al}. 1998).  As with 
Q0957+561 G1, we model this galaxy by including constraints on $I(R)$ and 
the central $\hat{\sigma}_{\rm p}$ along with the mean profile constraints.
We find that these constraints are not sufficient to break the degeneracy
between the different lens models, so that a large range of values is still
permitted for $H_0$ (44-68 km s$^{-1}$ Mpc$^{-1}$).
Further constraints on the mass distribution in this system are needed.

The stellar dynamics of gravitational lens systems show considerable 
promise for determining $H_0$, even if we have only central dispersions.
Larger samples of dispersion measurements would be particularly valuable,
since each lens system will have a different set of systematic 
uncertainties
(it is encouraging to note that the results for $H_0$ from Q0957+561 and
PG 1115+080 are so far consistent with each other).
Of particular value would be results from isolated systems 
with lens geometries that probe the radial mass distribution, such as 
MG 1654+1346 (see Ellithorpe, Kochanek, \& Hewitt 1996) and MG 1549+3047 
(Leh\'ar {\it et al}. 1993, 1996).  The results from a large set of these 
systems could be combined using Bayesian methods to converge on a robust 
value for $H_0$ (see Press 1997), avoiding the correlated systematic 
uncertainties that plague ``distance ladder'' approaches.
Furthermore, the 
independent measurement of galaxy properties from gravitational lensing and 
from stellar dynamics has an enormous potential for shedding light on the 
detailed structure of galaxies.

\vskip 67pt

We thank Ralf Bender and Roberto Saglia for providing their data in tabular 
form; Marijn Franx, Ramesh Narayan, John Tonry, and the anonymous referee for helpful comments;
and Roeland van der Marel for both.  We also thank Norm Grogin for many further 
demonstrations of his lens models.  C.S.K. is supported by NSF grant AST 
94-01722 and NASA ATP grant NAG 5-4062.

\newpage

\end{document}